# Limits on intrinsic magnetism in graphene

M. Sepioni[1], R. R. Nair[1], S. Rablen[1], J. Narayanan[1], F. Tuna[2], R. Winpenny[2], A. K. Geim[1]*, I.V. Grigorieva[1]*

[1]Manchester Centre for Mesoscience & Nanotechnology and [2]School of Chemistry, University of Manchester, Manchester M13 9PL, UK

*We have studied magnetization of graphene nanocrystals obtained by sonic exfoliation of graphite. No ferromagnetism is detected at any temperature down to 2 K. Neither do we find strong paramagnetism expected due to the massive amount of edge defects. Rather, graphene is strongly diamagnetic, similar to graphite. Our nanocrystals exhibit only a weak paramagnetic contribution noticeable below 50K. The measurements yield a single species of defects responsible for the paramagnetism, with approximately one magnetic moment per typical graphene crystallite.*

The long-standing interest in magnetic behavior of pure carbon-based systems has been further stimulated by reports of room-temperature ($T$) magnetic ordering in highly oriented pyrolytic graphite (HOPG) [1], nanographites [2], nanodiamonds [3] and disordered carbon films [4]. Although in these studies magnetization signals $M$ were small (typically, less than ~0.1 emu/g, i.e., less than 0.1% of the magnetization of iron), a consensus is emerging that, despite the absence of *d*- or *f*- electrons, magnetism in carbon systems may exist under a variety of experimental conditions. Furthermore, it is shown theoretically that atomic scale defects in graphene-based materials, e.g. adatoms and vacancies, can carry a magnetic moment $\mu$ of about one Bohr magneton, $\mu_B$ [5-8]. Also, extended defects such as edges can give rise to $M$ [9,10]. The possibility of long-range magnetic ordering has been predicted for randomly distributed point defects and grain boundaries [6,8], and bilayer graphene was suggested to exhibit spontaneous many-body ferromagnetism [11]. All this leaves little doubt that magnetism in graphene-based systems can in principle exist, although the whole subject remains highly controversial, especially as concerns (i) the role of environment and magnetic contamination [12] and (ii) the mechanism that could lead to the strong interaction required for ferromagnetism at room $T$.

The recent interest in isolated graphene has inevitably led to the question of possible ferromagnetism in this novel material too, especially due to the fact that it presents the basic structural element for all other graphitic forms [13]. The first experiments reported room-$T$ ferromagnetism in bulk samples obtained by conversion of nanodiamond and arc evaporation of graphite [14] and in graphene oxide [15]. In both studies, magnetic signals were again small (saturation magnetization $M_S$~0.1-1 emu/g) and have left open the same questions that haunt the previous reports of room-$T$ ferromagnetism in carbon materials. The dimensionality of graphene makes it even harder to explain the ferromagnetism theoretically.

In this Letter, we have studied magnetization of graphene obtained by direct ultrasonic cleavage of high-purity HOPG [16]. The resulting samples were laminates consisting of mostly mono- and bi-layer crystallites with typical sizes of 10 to 50 nm, aligned parallel to each other and rotationally disordered. The samples weighed several mg and were suitable for SQUID magnetometry. We found that the laminates are strongly diamagnetic and exhibit no sign of ferromagnetism at any $T$. Only by employing fields $H$ up to 70 kOe, we have detected a notable low-$T$ paramagnetic contribution ($M_S \approx 0.1$ emu/g). The paramagnetism is orders of magnitude smaller than that expected for the large number of broken bonds present in the laminates. By varying preparation procedures and environmental factors, we found that the paramagnetism is rather reproducible and, in further control experiments including the use of X-ray fluorescence spectroscopy (XRFS) and boron nitride laminates, ruled out any contamination with magnetic impurities.

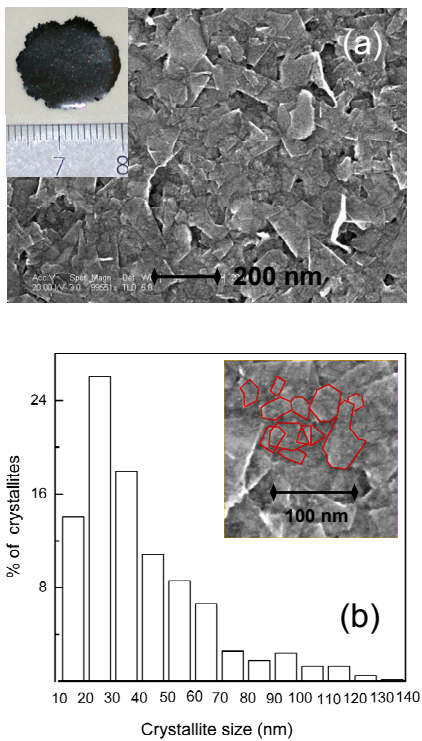

FIG. 1. Graphene laminates. (a) Typical SEM micrograph. Inset: photo of the whole sample. (b) Histogram shows the size distribution for 300 crystallites found within the ~1μm² area imaged in (a). Crystallite sizes were determined as geometrical averages. The inset zooms into the central region of (a). Edges of some of the smallest crystals are outlined for clarity.

Our samples (Fig. 1) were prepared by following the procedures reported in [16]. In brief, HOPG crystals – the cleanest form of graphite available (XRFS shows no paramagnetic impurities at a level of 1ppm) – were exfoliated by extensive sonication, using different organic solvents, namely, chloroform, dimethylformamide (DMF) and N-methylpyrrolidone. The suspensions were centrifuged to obtain stable solutions. These were passed through alumina filters, which resulted in the deposition of graphene crystallites forming several μm thick laminates (Fig. 1a). Extreme care was taken to use highest purity solvents (content of magnetic impurities <1ppm) and also to avoid any contamination. To verify the impurity content in the laminates we employed XRFS and found no $f$ or $d$ impurities above a detection limit of ≈10ppm [17].

Fig. 1 shows a typical scanning electron microscopy (SEM) image of our samples. One can see that individual crystallites mostly have irregular shapes but the edges follow main crystallographic directions [13] (Fig. 1b). The smallest crystals tend to have a distorted hexagon shapes. The majority of crystals are very small, with 60% having sizes below 40 nm, that is, much smaller than the sizes reported for short or mild sonication [16]. The histogram shown in Fig. 1b is characteristic for all our samples (for details, see [17]). Previous studies of similar suspensions showed that ~30% of crystallites were monolayers, with the rest made up of 2 to 5 layers [16,18]. By using transmission electron microscopy (TEM), we counted crystals of different thicknesses and found that our samples contained a larger proportion of monolayers (up to 50%), apparently due to more extensive sonication. The separation between graphene planes in the laminates was analyzed by X-ray diffraction. The most frequently found spacing was ≈3.36 Å, with a further large proportion ranging from ≈3.37 to 3.86 Å, i.e., significantly larger that the interlayer distance in graphite (3.334 Å). This, together with the rotational disorder (as seen by SEM and TEM), implies that during the filtration crystals do not restack and register into graphite but form a collection of electronically decoupled nanocrystals [19].

Magnetization measurements were performed using a SQUID magnetometer MPMS XL7. HOPG exhibited room-$T$ diamagnetic mass susceptibility $\chi_m \approx -3 \cdot 10^{-5}$ emu/g (dimensionless c.g.s. susceptibility $\chi \approx -6.5 \cdot 10^{-5}$) in $H$ perpendicular to graphene and $\chi \approx -8.5 \cdot 10^{-7}$ in parallel $H$, in agreement with literature values. The diamagnetism slightly increased as $T$ decreased from 300 to 100 K and became essentially $T$ independent at lower $T$. No paramagnetism was detected in HOPG at any $T$ within our experimental accuracy. Similar to HOPG, graphene laminates exhibited strong but distinctly smaller diamagnetism: $\chi \approx -1.5 \cdot 10^{-5}$ in perpendicular $H$. In parallel $H$, laminates were somewhat more diamagnetic than HOPG (Fig. 2), which is attributed to crystallites being not perfectly aligned. No ferromagnetism was detected at any $T$. These observations are in stark disagreement with the reports of room-$T$ ferromagnetism in graphene-like materials [14,15] and, also, have implications for interpretation of the ferromagnetism observed in graphite and other graphitic materials.

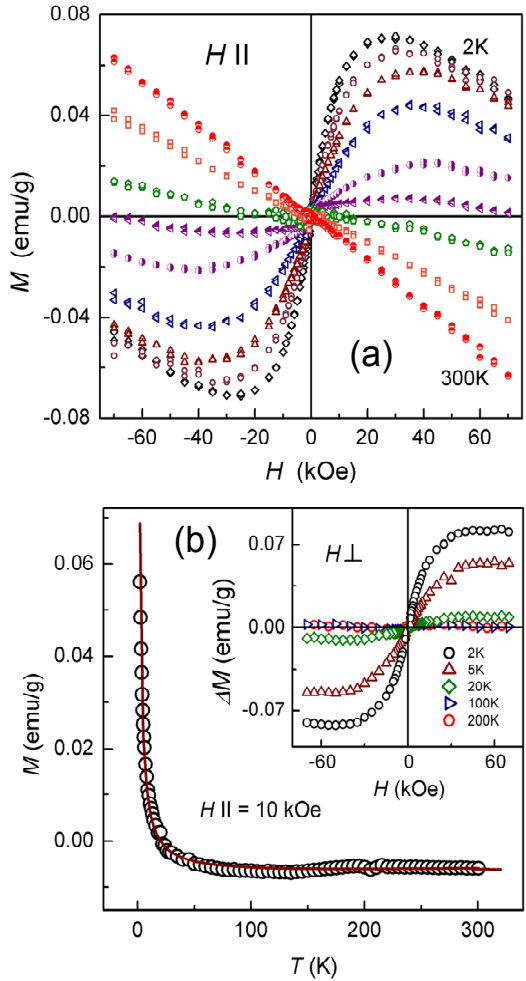

FIG. 2 (color online). Magnetic response of graphene. (a) Magnetic moment $M$ as a function of parallel $H$ at different $T$: (from top to bottom) 2, 3, 4, 5, 10, 15, 20, 50 and 300K. (b) $M(T)$ in parallel $H$ for the sample in (a). Symbols are the measurements; the curve is the Curie law calculated self-consistently (see text) with an account taken of a constant diamagnetic background ≈-0.008 emu/g in this particular $H$. Inset: Excess moment $\Delta M$ after subtracting the diamagnetic background (measured at room $T$) as a function of perpendicular $H$.

Despite the absence of ferromagnetism, our samples exhibited noticeable low $T$ paramagnetism, which is discussed in the rest of the paper. Fig. 2 plots the measured mass magnetization $M$ as a function of $H$ and $T$. One can see that as $T$ decreases below 20 K, the magnetization response in parallel $H$ becomes positive. As $T$ is lowered further, a typical paramagnetic behavior emerges, with low-field susceptibility $\chi=M/H$ following the Curie law $\chi \propto 1/T$ (Fig. 2b). In perpendicular $H$, magnetization was dominated by diamagnetism, as expected. Nevertheless, after subtracting the linear background, $\Delta M(H,T)$ curves showed exactly the same paramagnetic contribution as in parallel $H$ (Fig. 2b), i.e. the paramagnetism is isotropic. To characterize the magnetic species contributing to the observed behavior, we plot $M$ as a function of the reduced field $H/T$. Fig. 3 shows that all the $\Delta M(H/T)$ dependences collapse on a single curve, indicating a single type of non-interacting spins present in graphene. The observed behavior is well described by the standard Brillouin function:

$$M = NgJ\mu_B\left[\frac{2J+1}{2J}\text{ctnh}\left(\frac{(2J+1)x}{2J}\right) - \frac{1}{2J}\text{ctnh}\left(\frac{x}{2J}\right)\right]$$

where $x = gJ\mu_B H/k_B T$ and $k_B$ is the Boltzmann constant. The $g$ factor and the angular momentum number $J$ define the initial slope of $M(H/T)$ whereas the saturation level depends on the number of present spins, $N$. Assuming $g=2$, the Brillouin function provides excellent fits for $J=2$ and $5/2$ (Fig. 3). Self-consistently, the Curie law $M/H = [NJ(J+1)g^2\mu_B^2]/(3k_BT)$ with $J=5/2$ and $N=2.2\cdot10^{18}$ g$^{-1}$ inferred from Fig. 3 also gives an excellent fit to $M(T)$ dependence in Fig. 2b. $M(T)$ calculated for $J=2$ ($N=2.8\cdot10^{18}$g$^{-1}$) provides an equally good fit (not shown). If for some reasons the $g$ factor is enhanced, the experimental data can be described by a smaller $J$ but the fit becomes progressively poor, and only $J=3/2$ (that requires $g\approx2.5$) cannot be ruled out. The trivial free-electron $J=1/2$ expected for vacancies and most adatoms [5-8] cannot fit the data. Fig. 3 allows us to conclude that the observed paramagnetism is due to a single species with $\mu=gJ\mu_B \approx 4$-$5\mu_B$ and concentration $\approx50$ ppm (one moment per 20,000 carbon atoms or per 40x40 nm$^2$ crystal).

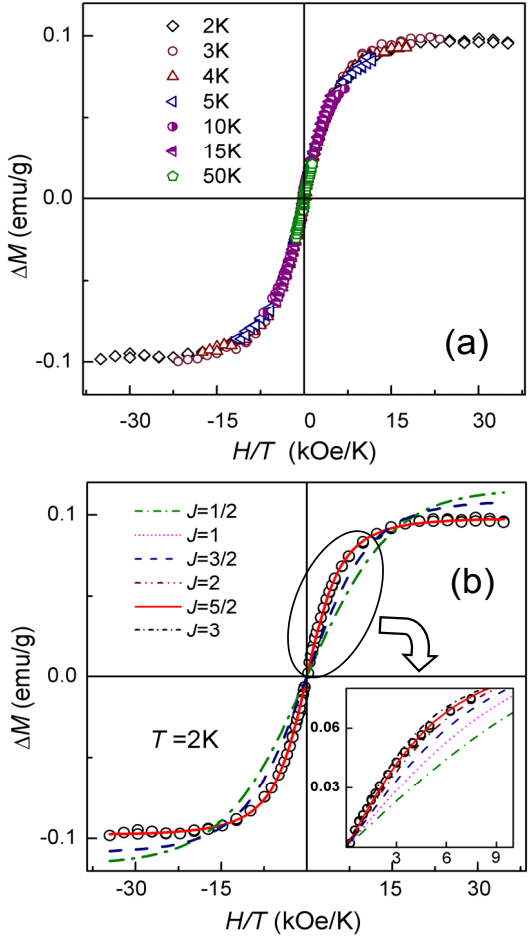

FIG. 3. Analysis of graphene's weak paramagnetism: (a) Magnetization curves of Fig. 2a plotted as a function of reduced field $H/T$ with the diamagnetic background subtracted. (b) Fits of the data in (a) using the Brillouin function with different values of $J$. For clarity, only data at 2K are used here. Inset: Zoom of the low-$H$ part of the graph, which is most sensitive to $J$.

The question that usually arises when ferromagnetism is reported for materials that contain no $f$ or $d$ electrons, is whether the observed signals can be explained by contamination. XRFS detected no paramagnetic impurities at a level of 10 ppm over the whole sample (cf. [4,14,15]). Nonetheless, we cross-checked this conclusion in a complementary study where we intentionally allowed a small amount of paramagnetic contamination by using a standard grade DMF (~5 ppm Fe). As a result, XRFS detected $\approx20\pm5$ ppm of Fe in the resulting laminates [17] whereas SQUID measurements yielded an extra paramagnetic contribution of $\approx15\pm5$ ppm. The agreement between the XRFS and SQUID analyses proves that our XRFS was reliable in discerning a minute magnetic contamination. Its amount needed to produce the observed $M$ in clean laminates would be detected easily. In another control experiment, we used boron nitride (BN) to make similar laminates (the two materials have similar structural but not electronic properties). No para- or ferro- magnetism was detected in the BN laminates.

What could be the origin of the detected moments? Unlike room-$T$ ferromagnetism, intrinsic paramagnetism with $J =1/2$ would agree with the existing theories because vacancies, adatoms and edges can carry localized moments [5-10]. Typical levels of chemical doping in graphene are ~1,000 ppm ($10^{12}$ cm$^{-2}$) [13], and XRFS detected several nonmagnetic elements with concentrations reaching sometimes up to ≈200 ppm for Ca, 250 ppm for S and 1,000 ppm for Cl [17], depending on the used solvent. Some of these nonmagnetic impurities can in principle bind to graphene and generate magnetic moments [5-7]. However, the measured $M(H,T)$ were reproducible in different runs whereas concentrations of nonmagnetic impurities varied randomly (e.g., no Cl or S was detected in some samples). Also, both *in* and *ex situ* annealing at $T$ up to 600°C did not result in magnetization changes. This indicates that the observed paramagnetism is related to structural rather than chemical defects. Furthermore, we found a notable reduction in $M$ for laminates with larger crystallites (due to shorter sonication), although this can also be related to a larger portion of multilayers. Annealing in oxygen at 450°C, which etched holes in graphene [20], led to a notable (~50%) increase in $M$, which also points in the direction of edge-related magnetism. For our typical crystals, the number of broken bonds along the edges is a few %. If we assume that $\mu_B$ is associated with each non-bonding electron, the number of spins contributing to paramagnetism would be ~$10^4$ per million atoms, i.e. 2 orders of magnitude more than observed. This proves that most of the broken bonds do not contribute to magnetism, being reconstructed or passivated [9].

Magnetic moments in graphene can be associated not only with point defects but also with extended ones such as zigzag edges [9,10]. In this case, the magnetic moment would depend on the total length of zigzag segments and in principle can be arbitrarily large. At first glance, this mechanism seems to lack an explanation for the value of $M_S$ being much smaller than the available broken bonds could generate. However, a recent theory [21] suggests that, due to interactions between different zigzag segments in sub-100 nm samples of a random shape, just a small number of non-compensated spins can survive (<10), which depends on sample size only logarithmically. This is in agreement with our observation that the paramagnetism corresponds approximately to one magnetic moment per crystallite. However, we cannot exclude that the observed signal comes from bilayer or even trilayer nanocrystals whose electronic structure allows more options for the emergence of paramagnetism [11]. Our main conclusion is however the absence of any sign of ferromagnetism in graphene even at 2 K.

This work was supported by EPSRC, ONR, AFOSR and the Royal Society.